\documentclass[10pt,twocolumn,letterpaper]{article}

\usepackage{iccv}
\usepackage{times}
\usepackage{epsfig}
\usepackage{graphicx}
\usepackage{tabularx}
\usepackage{graphicx}
\usepackage{adjustbox}
\usepackage{amsmath}
\usepackage{amssymb}
\usepackage[dvipsnames]{xcolor}
\usepackage{duckuments}
\usepackage{multirow}
\usepackage{booktabs}
\usepackage{authblk}

\usepackage[leftcaption]{sidecap}

\def\eg{\emph{e.g}\onedot} 
\def\ie{\emph{i.e}\onedot}

\def\etal{\emph{et al}\onedot}

\newcommand{\fref}[1]{Fig.~\ref{#1}}
\newcommand{\tref}[1]{Table~\ref{#1}}

\newcommand{\aref}[1]{Please refer to the supplementary materials}

\newcommand{\NTIRE}{NTIRE$_{syn}$}
\newcommand{\AIM}{AIM$_{syn}$}
\newcommand{\DPED}{DPED$_{rw}$}
\newcommand{\FACES}{FACES$_{rw}$}

\makeatletter
\newcommand{\printfnsymbol}[1]{%
  \textsuperscript{\@fnsymbol{#1}}%
}

\usepackage[pagebackref=true,breaklinks=true,letterpaper=true,colorlinks,bookmarks=false]{hyperref}

\iccvfinalcopy 

\def\iccvPaperID{13} 

\ificcvfinal\fi

\begin{document}

\title{Generalized Real-World Super-Resolution through Adversarial Robustness} 

\author[1]{Angela Castillo\thanks{equal contribution}}
\author[1]{María Escobar\printfnsymbol{1}}
\author[1,2]{Juan C. Pérez}
\author[3]{Andrés Romero}
\author[3]{Radu Timofte}
\author[3]{Luc Van Gool}
\author[1]{Pablo Arbeláez}

\affil[1]{Center for Research and Formation in Artificial Intelligence, Universidad de los Andes, Colombia}
\affil[2]{King Abdullah University of Science and Technology (KAUST), Saudi Arabia}
\affil[3]{Computer Vision Lab, ETH Zürich, Switzerland}
\affil[]{{\tt\small \{a.castillo13, mc.escobar11, jc.perez13, pa.arbelaez\}@uniandes.edu.co}

$^3${\tt\small roandres@ethz.ch, \{radu.timofte, vangool\}@vision.ee.ethz.ch}}


\maketitle
\ificcvfinal\thispagestyle{empty}\fi

\begin{abstract}
Real-world Super-Resolution (SR) has been traditionally tackled by first learning a specific degradation model that resembles the noise and corruption artifacts in low-resolution imagery. Thus, current methods lack generalization and lose their accuracy when tested on unseen types of corruption. In contrast to the traditional proposal, we present Robust Super-Resolution (RSR), a method that leverages the generalization capability of adversarial attacks to tackle real-world SR. Our novel framework poses a paradigm shift in the development of real-world SR methods. Instead of learning a dataset-specific degradation, we employ adversarial attacks to create difficult examples that target the model's weaknesses. Afterward, we use these adversarial examples during training to improve our model's capacity to process noisy inputs. We perform extensive experimentation on synthetic and real-world images and empirically demonstrate that our RSR method generalizes well across datasets without re-training for specific noise priors. By using a \emph{single} robust model, we outperform state-of-the-art specialized methods on real-world benchmarks. 
\end{abstract}

\section{Introduction}

Super-Resolution (SR) is the task of increasing the resolution of a given image. 
The ever-growing use of deep learning has fostered the creation of SR models that obtain remarkable results with high fidelity on traditional SR benchmarks~\cite{ledig2017photo,lim2017enhanced,lugmayr2020srflow,shocher2018zero,wang2018esrgan,zhang2019ranksrgan}. These conventional models are trained in a supervised manner with a High-Resolution (HR) and a corresponding Low-Resolution (LR) image pair. Since capturing the exact same scene in both HR and LR is complex and time-consuming, traditional SR datasets use clean HR images and the LR images are usually generated through a bicubic down-sampling operation~\cite{agustsson2017ntire,Timofte_2017_CVPR_Workshops}. Using a bicubic kernel on clean HR images simplifies the ill-posed SR task because it ignores the fact that real-world images are subjected to sensor noise and artifacts. Thus, models trained with clean, paired datasets tend to underperform when evaluated on real-life scenes.

\begin{figure}[t]
\begin{center}
\includegraphics[width=0.48\textwidth]{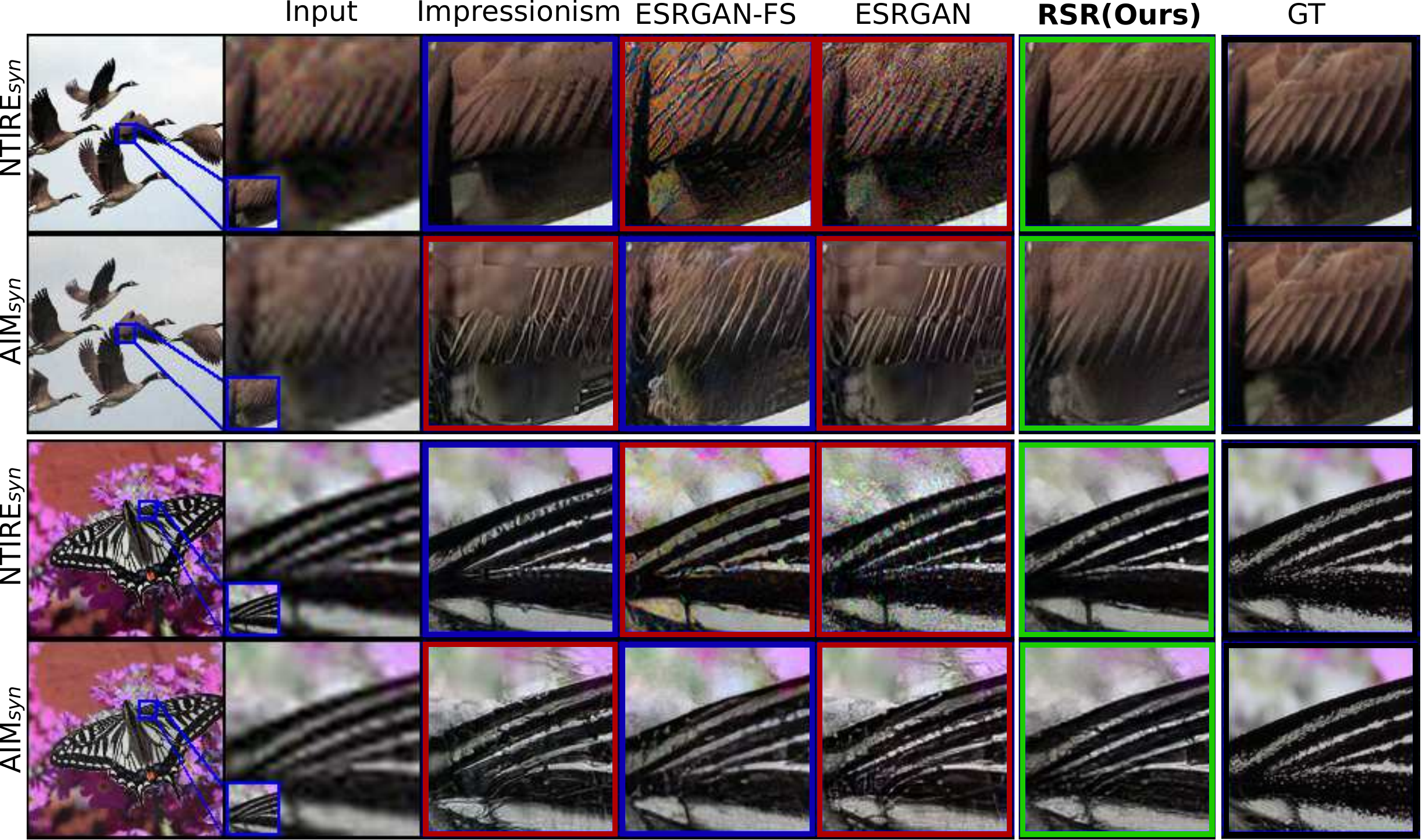}
\end{center}

   \caption{We present the comparison between our method and state-of-the-art methods: Impressionism~\cite{ji2020real} and ESRGAN-FS~\cite{fritsche2019frequency}, for two different types of degradations: \NTIRE~\cite{lugmayr2020ntire} and \AIM~\cite{lugmayr2019aim}. For reference, we show the bicubically upsampled input, the result of a supervised SISR method (ESRGAN~\cite{wang2018esrgan}), and the ground-truth (GT). {\color{blue}Blue} frames denote training and validation on the same dataset. {\color{red}Red} frames denote training and validation on different datasets. {\color{green}Green} frames denote our method.}
\label{fig:fig1}
\end{figure}

This limitation motivated the study of real-world SR on datasets with synthetic and natural corruptions \cite{lugmayr2020ntire,lugmayr2019aim}. Several benchmarks design real-world artifacts and corruptions under different assumptions or from varying sensors, and current solutions have to produce a specific model for each one.
Consequently, state-of-the-art models for real-world SR~\cite{fritsche2019frequency,ji2020real} generate photo-realistic results only when they are evaluated with the particular dataset for which they were trained, failing to generalize to new datasets with unseen corruptions. As an example, \fref{fig:fig1} shows two of the best methods in real-world SR, Impressionism~\cite{ji2020real} and ESRGAN-FS~\cite{fritsche2019frequency}, evaluated on two benchmark datasets with different artificial corruptions. When evaluated on the image from the same training dataset (blue frames), both methods show excellent performance. However, they create significant artifacts in the image from the unseen dataset (red frames). 

Even though the different noise in each of the input images does not modify human perceptual recognition (\ie, we can easily discriminate the introduced artifacts in both cases), state-of-the-art models are not \textit{robust} to unseen noise. To the best of our knowledge, there is no single real-world SR method that can generalize across different types of noise.

A way to assess a model's robustness to unseen noise is through Adversarial Attacks \cite{madry2018towards}. These attacks are based on adversarial examples, which are small intensity perturbations to the input image specifically designed to trick the model into failure cases. Adversarial examples have demonstrated improvement in noise generalization of models for various tasks such as classification~\cite{madry2018towards,szegedy2014intriguing}, semantic segmentation~\cite{arnab2018robustness, cisse2017houdini,xie2017adversarial} and object detection~\cite{zhang2019detection}. Through robust training~\cite{madry2018towards}, models learn to be invariant to noise by creating more human-perception aligned filters \cite{engstrom2019adversarial}. 

Choi \etal \cite{choi2019evaluating} explored the effect of adversarial attacks on traditional SR methods, finding that these methods are brittle and easy to fool under adversarial examples. In more recent work, Choi \etal ~\cite{choi2020adversarially} proposed an adversarial defense by modifying the intermediate filters of the network, thus improving the performance of traditional SR models under their previous attacks. However, their work was focused on evaluating the methods based on pixel-wise metrics. To date, there are no works that report the study of robust training for the real-world SR problem. 

In this work, we leverage adversarial attacks to create a real-world SR model robust to unseen noise types. As illustrated in \fref{fig:method}b, we start with a standard SR model trained on a clean dataset and find adversarial examples of the LR input that damage the performance of the model. Simultaneously, we employ the created examples to teach the network that small perturbations in the LR image should result in the same HR ground-truth. Our approach presents a paradigm shift in the study of real-world SR because, instead of generating LR images with a type of noise that resembles that of a specific dataset, we focus on making the SR model more robust to any input. \fref{fig:fig1} shows the generalization capability of our method. Even though our model has never seen images from either of the artificially corrupt datasets, by applying robust training, we are able to obtain photo-realistic HR images for both types of input noise.

We perform extensive quantitative and qualitative evaluations on data from the NTIRE 2020 Challenge on Real-World Image Super-Resolution~\cite{lugmayr2020ntire} and the AIM 2019 Real World Super-Resolution Challenge~\cite{lugmayr2019aim}. We show that our \emph{single} model, trained robustly on clean images, achieves state-of-the-art performance compared to methods with specialized models for each specific dataset. We also provide qualitative evaluation and non-reference perceptual metrics for images retrieved from an iPhone 3~\cite{ignatov2017dslr} and  facial super-resolution in the wild~\cite{yang2016wider}. 

Our main contribution is twofold: first, we propose a novel use of adversarial attacks in real-world super-resolution. Second, through the use of adversarial examples, we are able to create a generalized real-world SR model that achieves state-of-the-art results without training or fine-tuning on corrupt or real-world datasets. The code for reproducing our results is available at \href{https://github.com/BCV-Uniandes/RSR}{https://github.com/BCV-Uniandes/RSR}. 

\section{Related work} 

\subsection{Image Super-Resolution}
\subsubsection{Single Image Super-Resolution}
Single Image Super-Resolution (SISR) is a problem that has been widely studied in computer vision \cite{chowdhuri2012very, pan2018learning,yang2007spatial,guo2020closed}. Early approaches incorporated simple linear interpolation methods \cite{keys1981cubic,zhang2006edge} estimating the correlation \cite{allebach1996edge} or covariance of the low-resolution data \cite{li2001new}. Nevertheless, those methods fell short to fully capture the high frequencies of the images.

More recent methods have focused on improving the model's architecture to have better performance, either optimizing for pixel-wise scores (\eg, PSNR-oriented) or GAN-based~\cite{goodfellow2014generative} perceptual scores (\eg, LPIPS-oriented~\cite{zhang2018unreasonable}). SISR is very popular~\cite{ledig2017photo,lim2017enhanced,lugmayr2020srflow,shocher2018zero,wang2018esrgan,zhang2019ranksrgan} in its simplified form, where it is relatively easy to create paired datasets using a known downsampling strategy (\eg, bicubic interpolation). However, assuming prior knowledge of the downsampling kernel is unrealistic for images in the wild, as the kernel and the type of degradation are unknown, and therefore, SISR models fail to generalize. Moreover, images in the wild might contain corruptions and artifacts that should be removed in the super-resolution process. Therefore, blind SR approaches~\cite{bell2019blind,michaeli2013nonparametric,gu2019blind,zhang2020deep,Ren_2020_CVPR} tackle unknown degradation operations, whereas unsupervised real-world approaches~\cite{lugmayr2019unsupervised,maeda2020unpaired,bulat2018learn,yuan2018unsupervised,kim2020unsupervised,ren2020real,zhou2020guided} tackle severe degradation or compression artifacts in low-resolution images. Since our approach aims at performing super-resolution robust to noise and artifacts, we focus on the real-world literature.

\begin{figure*}[t]
  \centering
  \includegraphics[width=0.95\linewidth]{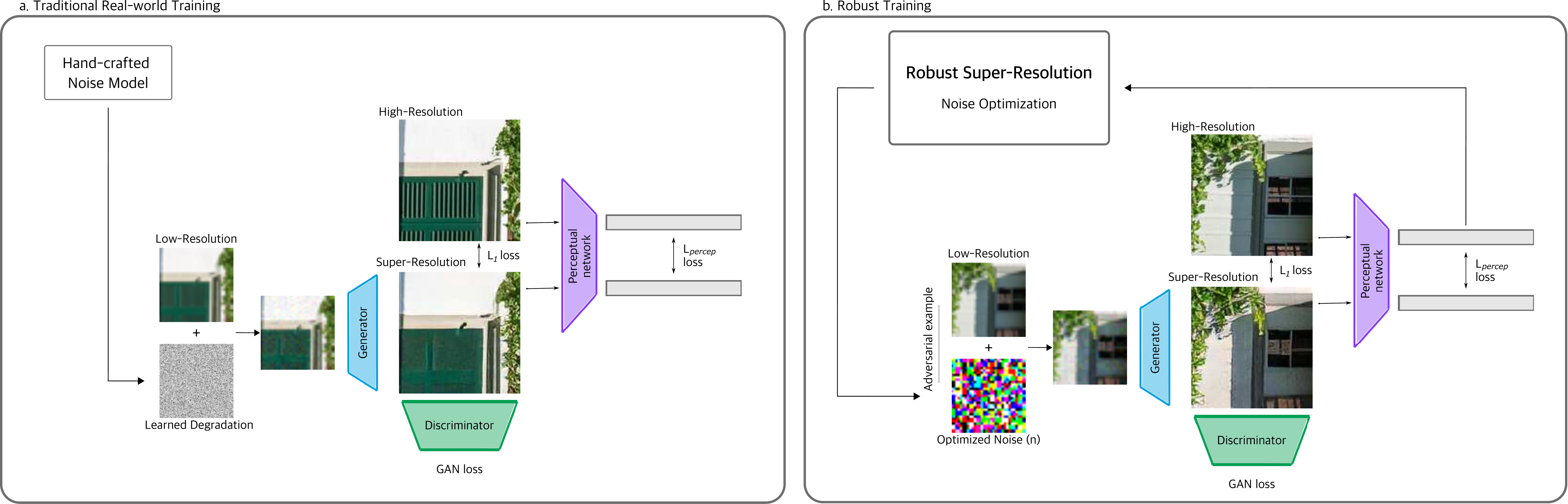}
  \caption{\textbf{Traditional \textit{vs.} Robust Super-Resolution.} In  \textit{(a) Traditional Real-world Training}, methods use hand-crafted noise models to learn dataset-specific degradations. In contrast, our  \textit{(b) Robust Training} strategy employs optimized adversarial examples to train a single SR model robustly. }
  \label{fig:method}
\end{figure*}

\subsubsection{Real-World Super-Resolution}
Real-world images contain noise or corruptions inherent to the sensor acquisition, or data compression artifacts. Traditional SR techniques~\cite{wang2018esrgan,zhang2019ranksrgan,ledig2017photo} fail at this task because they enhance the artifacts. Due to the problem formulation, it is hard to get paired real-world low-resolution and high-resolution alignment, so this problem is traditionally tackled in an unsupervised fashion by first learning a degradation procedure that can be used to produce an artificially paired dataset. As using paired datasets has been widely studied by SISR approaches, real-world methods focus on the degradation stage, that is, finding the best way to inject noise that resembles the one in the low-resolution dataset (\fref{fig:method}a).

\subsection{Adversarial Attacks}
Adversarial attacks on Deep Neural Networks (DNNs) were firstly conducted by Szegedy \etal \cite{szegedy2013intriguing}. 
Their results showed that DNNs could be fooled by imperceptible perturbations that do not modify the image semantics. Particularly, Choi \etal~\cite{choi2019evaluating, choi2020adversarially} demonstrated that SR methods are susceptible to adversarial attacks.
Formally, an adversarial attack is a procedure that, given an image, produces a poisoned input image that the DNN misclassifies with high confidence~\cite{goodfellow2014explaining}. Large efforts have been devoted to developing models that are both accurate \textit{and} robust to attacks. Adversarial Training (AT)~\cite{madry2018towards}, a technique that has stood the test of time, 
casts the problem of training adversarially-robust models as one of robust optimization of a saddle-point problem. In practice, AT trains a DNN on adversarial examples generated on-the-fly through Projected Gradient Descent (PGD) by aiming at maximizing the DNN's loss. 

While much of the initial concern regarding the brittleness of DNNs against attacks was related to security concerns, recent evidence provides new perspectives. Stutz \etal~\cite{Stutz2019CVPR} explored how adversarial examples may be related to generalization in learning algorithms. In relation to perception, Engstrom \etal \cite{engstrom2019adversarial} demonstrated that robust features are aligned with human perception, and hence adversarial robustness can serve as a prior for learned representations. Further, Santurkar \etal~\cite{santurkar2019image} showed that adversarially-robust classifiers prove to be useful for image synthesis tasks. Inspired by these findings, in this paper we demonstrate how adversarial robustness can be used as a useful prior for learning DNNs for real-world SR. In particular, we show that by using adversarial attacks we can bypass the need of corrupted low-resolution data, and train the robust super-resolution architecture directly, thus producing a \emph{single} model that generalizes to several datasets. 

To avoid ambiguity with Generative Adversarial Networks (GANs), we refer to Adversarial Training as Robust Training.
\section{Method}

Traditionally, real-world SR approaches design hand-crafted noise models for learning dataset-specific degradation and using it to create new LR images to train their SR models (\fref{fig:method}a). However, these models tend not to generalize well to new datasets with an unknown degradation distribution. Unlike traditional approaches, our method creates LR images that are perceptually challenging for the SR model in order to improve its generalization capability (\fref{fig:method}b). 

Inspired by recent works that have drawn a link between adversarial robustness and both generalization and human perception~\cite{engstrom2019adversarial, santurkar2019image}, we exploit adversarial robustness as useful prior knowledge for learning SR models. We underscore that, as with any ill-posed problem such as SR~\cite{yang2019deep}, useful priors are of utmost importance: what may prove to be a reasonable output from a SR model strongly relies on vast amounts of prior information humans possess about the real world. Thus, in search of SR models with improved performance on unseen types of degradation, we leverage adversarially-robust training to induce priors correlated with human perception. In particular, we encourage the training procedure to prefer robust solutions against adversarial attacks during the learning of SR models.

\begin{figure}[t]
\begin{center}
\includegraphics[width=0.35\textwidth]{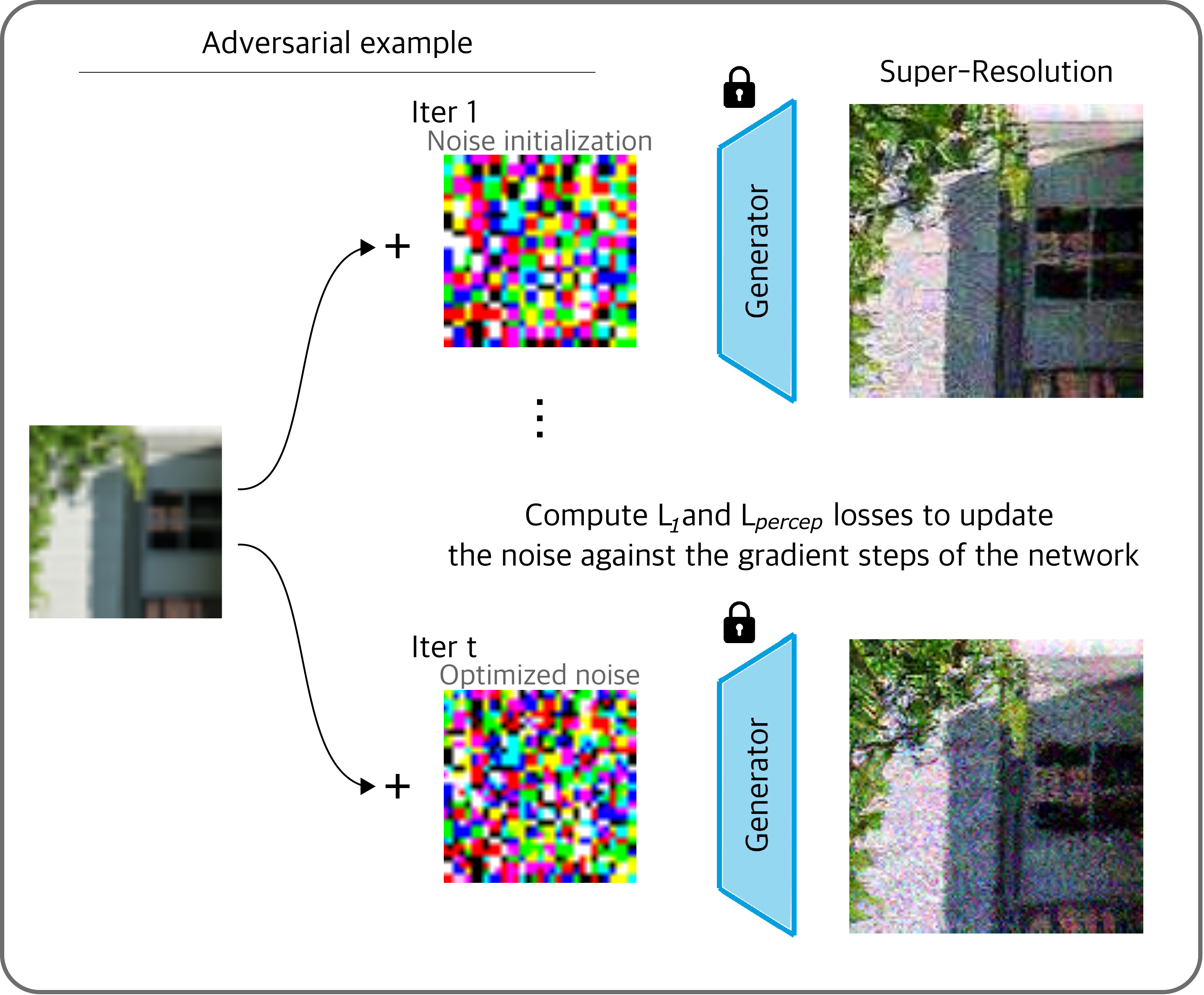}
\end{center}
  \caption{\textbf{Noise Optimization.} We create adversarial examples by finding iteratively a noise that exploits the weaknesses of the generator. Once the $t$ iterations of the optimization process are complete, the adversarial examples are used to improve our SR model during our \textit{Robust training}.}
  \label{fig:optim}
\end{figure}

\subsection{Mathematical Formulation}
\textbf{Super Resolution.} Let $I_{HR}$ be a HR image, and $I_{LR}$ be a LR image downsampled $\downarrow_{s}$ by a factor of $s$ as 
\begin{equation}
    I_{LR} = (I_{HR}\otimes k) \downarrow_{s} +~n,
\end{equation}
where $k$ is the downsampling kernel, and $n$ represents additive noise. In common SR methods~\cite{wang2018esrgan,ledig2017photo}, this formulation is simplified by using a bicubic kernel and small independent identically  distributed (\textit{i.i.d}) additive Gaussian noise~\cite{zhao2016fast}. In real-world approaches, $n$ is not neglected, as the corruptions severely alter the low-resolution image. To circumvent this issue, we optimize $n$ for each image using adversarial attacks. Without loss of generality, our model can be extended to blind approaches by considering unknown kernels $k$.

Our rationale to include adversarial examples that resemble real-world LR images is as follows: Employing GAN-based methods that include a generator to inject noise~\cite{bulat2018learn,lugmayr2019unsupervised} introduces a domain shift that compromises the performance during inference. Hand-crafted methods~\cite{ji2020real,fritsche2019frequency} do not introduce a domain shift because of their simplicity (\fref{fig:method}a) but require extensive manual labor to simulate different types of corruption. By using adversarial attacks, we can modify the injected noise and thus optimize it to find the hardest type of noise in a data-driven way (\fref{fig:method}b) while maintaining the semantic information present in an image. With this approach, we avoid a domain shift and present an optimized method to faithfully remove noise and corruption robustly.

\textbf{Robust Training.} In a fashion similar to~\cite{madry2018towards}, we generate adversarial examples on-the-fly via the Projected Gradient Descent (PGD) method. PGD is an optimization procedure for finding $\ell_{p}$-bounded adversarial examples. In this work, we focus in particular on $\ell_{\infty}$-bounded examples, computed as repeated iterations of 
\begin{equation} \label{eq:PGD}
    x^{adv}_{t+1} = \prod_{X} \Big(x_{t} + \alpha\: \text{sign}\big(\nabla_{x_{t}}L(x_{t},y)\big)\Big),
\end{equation}
where $\prod$ is the projection operator, $x_{t+1}$ is the input image of the attack iteration $t+1$, $y$ is the ground-truth, $X$ denotes the set of \textit{valid} images (\ie, the intersection of the $\ell_{\infty}$-ball of $\epsilon$ radius and the usual $[0,1]$ range for image pixels), $\alpha$ is the size of the gradient step and $L(\cdot)$ is the cost function with respect to which optimization is performed. For the first iteration, $x_0$ is perturbed with random noise of magnitude $\epsilon$ and projected back into $X$, as customary in previous works \cite{goodfellow2014explaining, kurakin2016adversarial}.
In our method, we use PGD to maximize both the $L_1$ and perceptual ($L_{percep}$) losses of the generator. Furthermore, including adversarial attacks within the SR framework requires significantly more ingenuity than just replacing the loss function. Specifically, besides setting an objective function for the target module, designing adversarial attacks during training requires defining \textit{(i)} how to maximize such objective, \textit{(ii)} the constraints to maximize the optimization variables, and \textit{(iii)} when/where to introduce adversarial examples. 

\subsection{Robust Training on Super-Resolution}
The underlying idea of robust training is to continually include adversarial examples during the training stage of the model. Our Robust Super-Resolution (RSR) model extends this idea to SR through three main steps. Firstly, we implement a GAN-based model for SR that is previously trained on clean LR images. Secondly, we employ the trained generator to create adversarial examples through a noise optimization process (\fref{fig:optim}). Finally, we use the adversarial examples as LR inputs to further train the GAN robustly (\fref{fig:method}b).

\textbf{GAN-based model.} We build upon ESRGAN~\cite{wang2018esrgan, wang2020basicsr}, a GAN-based approach to SR~\cite{goodfellow2014generative}. The Generator ($G$) is built upon the RRDBNet architecture~\cite{ledig2017photo} with Residual-in-Residual Dense Blocks to improve the enhanced image's quality. $G$ optimizes three losses: \textit{(i)} the $L_{1}$ loss to evaluate the pixel distance of the super-resolved image and the HR, \textit{(ii)} the $L_{percep}$ \cite{johnson2016perceptual} which is calculated as the distance of activation features of a pre-trained network between super-resolved and HR images, \textit{(iii)} and the term of the adversarial loss for the generator $L_{G}^{GAN}$. $L_1$ and $L_{percep}$ are minimized to reduce the distance between the SR and HR images. 
The objective function for $G$ is:
\begin{equation}
    L_G = L_{1} + L_{percep} + L_{G}^{GAN}.
\end{equation}

The Discriminator ($D$) is based on a VGG-128 architecture and follows the same principle as in Relativistic GAN~\cite{jolicoeur2018relativistic}, which estimates the probability that a real image is more realistic than a fake image. During the first stage of our method, we use an ESRGAN model pre-trained on clean images, \ie, the model can super-resolve clean images at this stage but it is not robust to any type of noise. 

\textbf{Noise Optimization.}
We use the pre-trained network $G$ as the starting point for our noise optimization process. Our goal is to find the optimal input noise that, when added to a LR image, results in a deficient super-resolved image. This process allows us to identify the generator's weaknesses in handling inputs with different types of corruption. \fref{fig:optim} depicts our optimization process. We freeze the weights of $G$ during this stage and create the adversarial examples based on the $L_1$ and $L_{percep}$ losses. We add an $\epsilon$-constraint random noise to the LR input image to retrieve an adversarial example.  Then, we optimize the noise through PGD (Eq. \ref{eq:PGD}). We iteratively update the input adversarial noise by considering the contribution of both losses and produce the adversarial example as follows:
\begin{equation}
\begin{aligned}
    I_{LR_{t+1}}^{adv} = I_{LR_{t}} +& \alpha\: \text{sign}\Big(\nabla_{I_{LR_{t}}}\big(L_1(G(I_{LR_{t}}),I_{HR}) \\
    +& L_{percep}(G(I_{LR_{t}}),I_{HR})\big)\Big).
\end{aligned}  
\end{equation}

The noise on real-world images tends to be grouped in multiple pixels, so we take inspiration from~\cite{cohen2019certified} to design structured noise by changing the input scale. To train our model, we start the optimization process with a synthetically grouped structure to better simulate real-world corruptions. Nevertheless, during the optimization iterations of the attack, we no longer control the structure of the noise as it changes according to the steps of the gradient. Please refer to the supplementary material for additional analysis regarding the noise structure.

\textbf{RSR Training.}
We use the adversarial examples created during the optimization stage as LR input to train the GAN further. Since the new inputs are specifically optimized to exploit the weaknesses of $G$, using them during training ensures that the network learns to super-resolve noisy images. Furthermore, since the optimized noise does not follow a pre-defined distribution, the network must generalize to different types of noise. The noise optimization and training stages are performed interspersed throughout the training iterations. Thus, as the training process of the network progresses, the examples that the network considers as adversarial are more potent after each new iteration. This process, in turn, makes the network more robust against increasing degradation in every new iteration.

\section{Experimental setup} 
\subsection{Datasets}

\textbf{Training.} The DIV2K dataset~\cite{agustsson2017ntire,Timofte_2017_CVPR_Workshops} is one of the widely used benchmarks for traditional SR. The training set contains 800 2K resolution images and their corresponding LR images, obtained using a bicubic downgrading operator. These images do not include any simulated or real noise, and we use them as the basis for our robust training. For memory-saving purposes during training, we crop the images into $480\times480$ sub-images.

\begin{table*}[]
\centering
\resizebox{\linewidth}{!}{%
\begin{tabular}{|cccc|c|c|c|c|c|c|c|c|c|}
\hline
\multirow{2}{*}{$\epsilon$} &
  \multirow{2}{*}{Iters} &
  \multirow{2}{*}{Noise structure} &
  \multirow{2}{*}{Loss} &
  \multicolumn{3}{c|}{PSNR$\uparrow$} &
  \multicolumn{3}{c|}{SSIM$\uparrow$} &
  \multicolumn{3}{c|}{\textbf{LPIPS}$\downarrow$} \\ \cline{5-13} 
   &   &   &   &  \NTIRE &  \AIM &  Avg &  \NTIRE &  \AIM &  Avg &
  \NTIRE &  \AIM &  Avg \\ \hline
\textbf{14/255} &
  \textbf{2} &
  \textbf{1.5} &
  \textbf{\underline{$L_{1}$ + $L_{percep}$}} &
  \textbf{24.31} &
  \textbf{21.99} &
  \textbf{23.15} &
  \textbf{0.65} &
  \textbf{0.60} &
  \textbf{0.62} &
  \textbf{{\color{red}0.23}} &
  \textbf{{\color{blue}0.37}} &
  \textbf{{\color{red}0.30}} \\
  14/255 &
  2 &
  1.5 &
  \underline{$L_{1}$} &
  {\color{red}25.62} &
  {\color{red}22.47} &
  {\color{red}24.04} &
  {\color{blue}0.68} &
  {\color{red}0.63} &
  {\color{blue}0.65} &
  0.30 &
  0.40 &
  0.35 \\
14/255 &
  2 &
  1.5 &
  \underline{$L_{percep}$} &
  24.20 &
  22.07 &
  23.13 &
  0.64 &
  0.61 &
  0.63 &
  {\color{blue}0.24} &
  0.38 &
  {\color{blue}0.31} \\ \hline
\underline{10/255} &
  2 &
  1.5 &
  $L_{1}$ + $L_{percep}$ &
  23.16 &
  21.84 &
  22.50 &
  0.56 &
  0.59 &
  0.58 &
  0.31 &
  {\color{blue}0.37} &
  0.34 \\
\underline{12/255} &
  2 &
  1.5 &
  $L_{1}$ + $L_{percep}$ &
  23.68 &
  21.74 &
  22.71 &
  0.62 &
  0.59 &
  0.60 &
  0.25 &
  {\color{blue}0.37} &
  0.31 \\
\underline{16/255} &
  2 &
  1.5 &
  $L_{1}$ + $L_{percep}$ &
  23.84 &
  22.06 &
  22.95 &
  0.63 &
  {\color{blue}0.62} &
  0.62 &
  0.25 &
  0.38 &
  {\color{blue}0.31} \\
\underline{18/255} &
  2 &
  1.5 &
  $L_{1}$ + $L_{percep}$ &
  24.02 &
  21.89 &
  22.96 &
  0.64 &
  0.61 &
  0.62 &
  0.25 &
  0.38 &
  {\color{blue}0.31} \\ \hline
14/255 &
  \underline{4} &
  1.5 &
  $L_{1}$ + $L_{percep}$ &
  24.13 &
  22.04 &
  23.09 &
  0.65 &
  0.61 &
  0.63 &
  0.26 &
  0.38 &
  0.32 \\
14/255 &
  \underline{6} &
  1.5 &
  $L_{1}$ + $L_{percep}$ &
  24.00 &
  21.73 &
  22.86 &
  0.63 &
  0.60 &
  0.62 &
  0.27 &
  {\color{blue}0.37} &
  0.32 \\
14/255 &
  \underline{8} &
  1.5 &
  $L_{1}$ + $L_{percep}$ &
  23.94 &
  21.56 &
  22.75 &
  0.63 &
  0.58 &
  0.61 &
  0.27 &
  {\color{red}0.36} &
  0.32 \\ \hline
14/255 &
  2 &
  \underline{1} &
  $L_{1}$ + $L_{percep}$ &
  {\color{blue}25.58} &
  {\color{blue}22.20} &
  {\color{blue}23.89} &
  {\color{red}0.70} &
  {\color{red}0.63} &
  {\color{red}0.67} &
  {\color{blue}0.24} &
  0.38 &
  {\color{blue}0.31} \\
14/255 &
  2 &
  \underline{2} &
  $L_{1}$ + $L_{percep}$ &
  19.99 &
  21.40 &
  20.69 &
  0.32 &
  0.54 &
  0.43 &
  0.54 &
  0.41 &
  0.47 \\ \hline
\end{tabular}%
}
\caption{\textbf{Ablation Study.} Performance comparison of varying the hyper-parameters for the noise optimization stage of our method. \emph{$\epsilon$} denotes the maximum pixel intensity perturbation of the noise. \emph{Iters} denotes the number iterations used to find the adversarial example. \emph{Noise structure} denotes how grouped the initial noise values are. \emph{Loss} indicates which loss function was set as objective for the noise optimization process. {\color{red}Red} and {\color{blue}blue} colors highlight the best two scores. The hyper-parameters we use for our RSR model are in \textbf{bold}.} 
\label{table:ablation}
\end{table*}

\textbf{Reference validation.} We validate our framework over the validation datasets from the NTIRE 2020 Challenge on Real-World Image Super-Resolution~\cite{lugmayr2020ntire} track 1, and the AIM 2019 Real World Super-Resolution Challenge~\cite{lugmayr2019aim} track 2. 
The LR images in each dataset include a different degradation obtained with an undisclosed artificial operator, resembling thus realistic corruptions and artifacts. The validation set is composed of the artificially degraded version of the 100 LR images from the DIV2K validation set and their corresponding HR ground-truth to calculate reference-based metrics. Since both datasets include synthetic corruptions, we refer to them as \NTIRE~and \AIM.

\textbf{Non-reference validation.} We also validate the performance of our method on real-world datasets that do not have a ground-truth HR for reference. We use the validation set of the DPED~\cite{ignatov2017dslr} dataset that contains real-world images taken by an iPhone3 camera. This dataset represents a more challenging framework because the data contains noise, differences in lighting, and other low-quality artifacts. Furthermore, we evaluate our method on real-world faces from the WIDER FACE dataset~\cite{yang2016wider}, initially designed for facial detection in-the-wild. We randomly sample 100 cropped faces of medium size with no occlusions. 
We refer to these datasets as \DPED~and \FACES. 

\subsection{Evaluation metrics}
As it is well known that pixel-wise metrics (\eg, PSNR) do not correlate well with human perception, the metric we focus on is the Learned Perceptual Image Patch Similarity (LPIPS)~\cite{zhang2018unreasonable}. This metric is based on the comparison between features of a neural network, in this case, a pre-trained AlexNet~\cite{krizhevsky2012imagenet} model, and it measures the distance in perceptual quality between the generated image and the ground truth. We also report the Peak Signal-to-Noise Ratio (PSNR) measured in decibels (dB) and Structural Similarity Index (SSIM) to evaluate the pixel-wise fidelity of the result with the ground-truth HR image. For the datasets that do not include ground-truth HR images, we use the non-reference image quality metrics: Perception-based Image Quality Evaluator (PIQE)~\cite{venkatanath2015blind}, Naturalness Image Quality Evaluator (NIQE)~\cite{mittal2012making}, and Blind/Referenceless Image Spatial Quality Evaluator (BRISQUE)~\cite{mittal2011blind}. We provide further validation 
in the supplementary material. 

\subsection{Implementation details}
\textbf{Robust training.} We retrieve the adversarial examples using PGD with parameter values $\epsilon$ = $14/255$, iterations = $2$, structured noise = $1.5\times$.  The ablation experiments for each of these parameters are presented in section \ref{sec:ablation}. We initialize training with the pre-trained weights of ESRGAN on DIV2K. Then, we perform our robust training on two NVIDIA QUADRO RTX 800 GPUs for 18k iterations with an initial learning rate of $1e^{-4}$, 16 images per batch, and use an Adam optimizer with $\beta_1 = 0.9$ and $\beta_2 = 0.99$. 

\textbf{Comparison with State-of-the-Art.} We compare our method against ESRGAN~\cite{wang2018esrgan} to establish a baseline against SISR models. We extend the ESRGAN framework for 18k iterations over the DIV2K dataset for a fair comparison. Furthermore, we compare our results against the best-performing method of the NTIRE20 and AIM19 challenges: Impressionism~\cite{ji2020real} and ESRGAN-FS~\cite{fritsche2019frequency}, respectively. 
For Impressionism, we use the provided pre-trained weights for \NTIRE~and \DPED, and for ESRGAN-FS, we use their weights for \AIM~and \DPED. For a complete comparison in every dataset, we train the available code of Impressionism in \AIM~and ESRGAN-FS in \NTIRE, using default parameters in their official public repositories. All our experiments are done considering a scaling factor of $4\times$.

\section{Results and Discussion}
This section analyzes the quantitative and qualitative results obtained by our method, and we perform a thorough comparison with the state-of-the-art methods. 

\subsection{Ablation study} \label{sec:ablation} 
Robust training relies on the preservation of the perceptual and semantic information of the original image. Creating hard examples for the network that are unrealistic compared to the training dataset would confuse the network rather than help encourage adversarial robustness. Thus, we have to find a trade-off between how hard the example is and how realistic it is. In \tref{table:ablation},  we present an extensive validation of the effect of each hyper-parameter to demonstrate that our final method uses the optimal one. Since our method aims to improve the perceptual quality, we focus mainly on LPIPS and report PSNR and SSIM as complementary metrics. Please refer to the supplementary material for qualitative visualizations and a more in-depth analysis of the different hyper-parameters.

\begin{figure*}[t]
  \centering
  \includegraphics[width=0.95\linewidth]{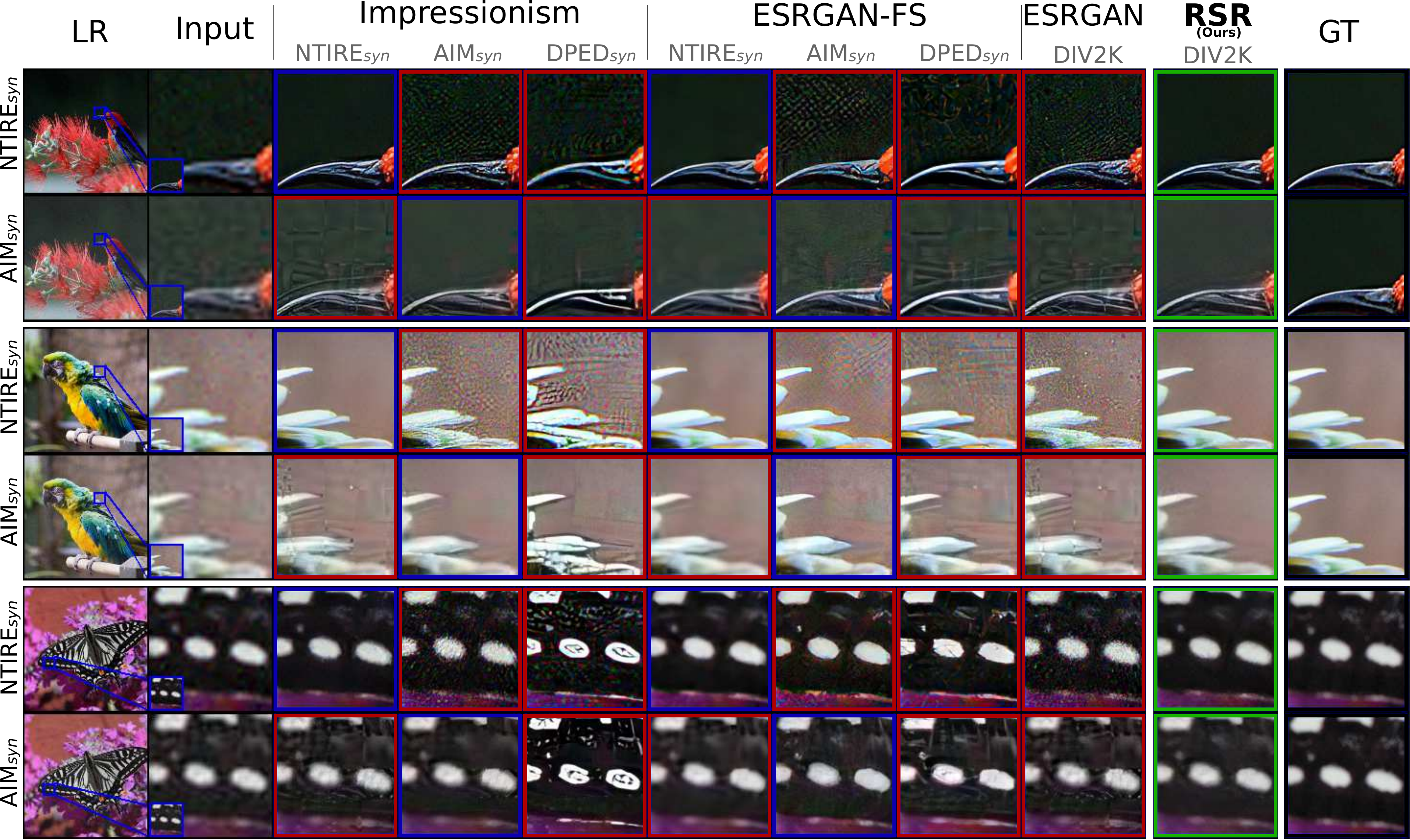}
  \caption{\textbf{Qualitative results on Synthetic Images.} Comparison between our method and state-of-the-art methods, for two synthetic corruption datasets: \NTIRE~and \AIM. For reference, we show the bicubically upsampled input, the result of a supervised SISR method (ESRGAN~\cite{wang2018esrgan}), and the ground-truth (GT). {\color{blue}Blue} frames denote training and validation on the same dataset. {\color{red}Red} frames denote training and validation on different datasets. {\color{green}Green} frames denote our method.}
  \label{fig:result_paired}
\end{figure*}

Choosing the right loss function to optimize is crucial because it directs the robust training towards creating hard examples for a part of the network. Using only the $L_1$ loss for robust adversarial optimization improves the pixel-wise metrics. This result is expected because we are specifically training the network to be more precise in a pixel-wise comparison. However, the perceptual quality is affected. In the same way, optimizing only the perceptual loss improves the LPIPS compared to the previous experiment. Nevertheless, when we use both losses, we are able to improve the average perceptual performance further. 

$\epsilon$ constrains the magnitude of the perturbation in the adversarial example. Larger values of $\epsilon$ result in larger perturbations as it allows the noise to be in a higher range, thus modifying the input more. \tref{table:ablation} shows that performing smaller modifications to the input, \ie, having a smaller $\epsilon$, greatly affects the mean LPIPS. However, the perceptual quality remains virtually the same when $\epsilon$ is large enough. 

Regarding the iterative nature of the attack, increasing the number of iterations leads to a worse result in every dimension of the evaluation. This result suggests that an adversarial attack with two iterations is strong enough to impact the SR model positively. 

Finally, the scale of the structured noise determines how grouped the initial noise is. Using a value of 1 means that every noisy pixel of the LR image has a different value. 
\tref{table:ablation} shows that a structured noise of 1 performs well for PSNR and SSIM but slightly worse for LPIPS. This phenomenon is the same that we find on the weakest $\epsilon$ modification. In contrast, the results suggest that using a value of 2 in the structure scale creates adversarial examples that are too different from the original input, so the network can not effectively use them for training. We find that the best trade-off is achieved with a structure scale of 1.5.  

\textbf{Effect of robust training.} \tref{tab:synthetic_images} shows that ESRGAN achieves the worst LPIPS for both \NTIRE~and \AIM. This behavior could result from training the model on clean images, depriving the model of seeing any corruptions like those present on the datasets. In contrast, RSR outperforms ESRGAN for every metric and has an average LPIPS 0.35 points lower. These results are visually confirmed in \fref{fig:result_paired}. ESRGAN magnifies the input noise of every image while RSR removes it. Since we use ESRGAN as our baseline, the only difference between that model and ours is that we perform robust training. Thus, even though we train with a clean dataset, the invariance to noise enforced during robust training significantly improves the generalization capacity of the model when evaluated on corrupt datasets.

\begin{SCtable*}[]
\resizebox{1.3\columnwidth}{!}
{%
\begin{tabular}{|c|l||c|c|c||c|c|c||c|c|c|}
\hline
\multicolumn{1}{|c|}{\multirow{2}{*}{Method}} & \multirow{2}{*}{\begin{tabular}[c]{@{}l@{}}Training\\ Dataset\end{tabular}} & \multicolumn{3}{c||}{\textbf{PSNR}$\uparrow$} & \multicolumn{3}{c||}{\textbf{SSIM}$\uparrow$} & \multicolumn{3}{c|}{\textbf{LPIPS}$\downarrow$} \\ \cline{3-11} 
\multicolumn{1}{|c|}{}                        &                                                                             & \NTIRE  & \AIM  & \textit{Avg}  & \NTIRE  & \AIM  & \textit{Avg}  & \NTIRE   & \AIM  & \textit{Avg}  \\ \hline
Bicubic                                       & -                                                                           & {\color{red}25.51}    & {\color{red}22.35}  & {\color{red}23.93} & {\color{blue}0.67}     & {\color{blue}0.62}   & {\color{blue}0.65}  & 0.63      & 0.68   & 0.66  \\ \hline
\multirow{3}{*}{Impressionism~\cite{ji2020real}}                & \NTIRE                                                                       & {\color{blue}24.82}    & 21.47  & 23.15 & 0.66     & 0.54   & 0.60  & {\color{red}0.23}      & 0.52   & 0.37  \\ 
                                              & \AIM                                                                         & 19.65    & 21.89  & 20.77 & 0.29     & 0.60   & 0.45  & 0.67      & 0.41   & 0.54  \\ 
                                              & \DPED                                                                        & 17.53    & 18.84  & 18.18 & 0.34     & 0.49   & 0.41  & 0.60      & 0.47   & 0.53  \\ \hline
\multirow{3}{*}{ESRGAN-FS~\cite{fritsche2019frequency}}                    & \NTIRE                                                                       & 24.59    & {\color{blue}22.07}  & {\color{blue}23.33} & {\color{red}0.69 }    & {\color{red}0.63}   & {\color{red}0.66}  & {\color{blue}0.25}      & 0.47   & {\color{blue}0.36}  \\ 
                                              & \AIM                                                                         & 19.56    & 20.82  & 20.19 & 0.31     & 0.51   & 0.41  & 0.56      & {\color{blue}0.39}   & 0.48  \\ 
                                              & \DPED                                                                        & 17.79    & 20.15  & 18.97 & 0.34     & 0.53   & 0.43  & 0.51      & 0.47   & 0.49  \\ \hline
ESRGAN~\cite{wang2018esrgan}                                      & DIV2K                                                                       & 20.59    & 21.48  & 21.03 & 0.43     & 0.56   & 0.49  & 0.68      & 0.53   & 0.60  \\ \hline
\textbf{RSR (Ours)}                                          & DIV2K                                                                       & 24.31    & 21.99  & 23.15 & 0.65     & 0.60   & 0.62  & \textbf{{\color{red}0.23}}      & \textbf{{\color{red}0.37}}   & \textbf{{\color{red}0.30}}  \\ \hline
\end{tabular}
}
\caption{\textbf{Quantitative comparison on Synthetic Images.} We report the comparison of reference metrics between our method and the state-of-the-art methods in different datasets. $\uparrow$ and $\downarrow$ indicate higher is better and lower is better, respectively. {\color{red}Red} and {\color{blue}blue} colors highlight the best two scores. \textbf{Bold} represents the best method for LPIPS metric.}
\label{tab:synthetic_images}
\end{SCtable*}

\begin{figure*}[t]
  \centering
  \includegraphics[width=0.9\linewidth]{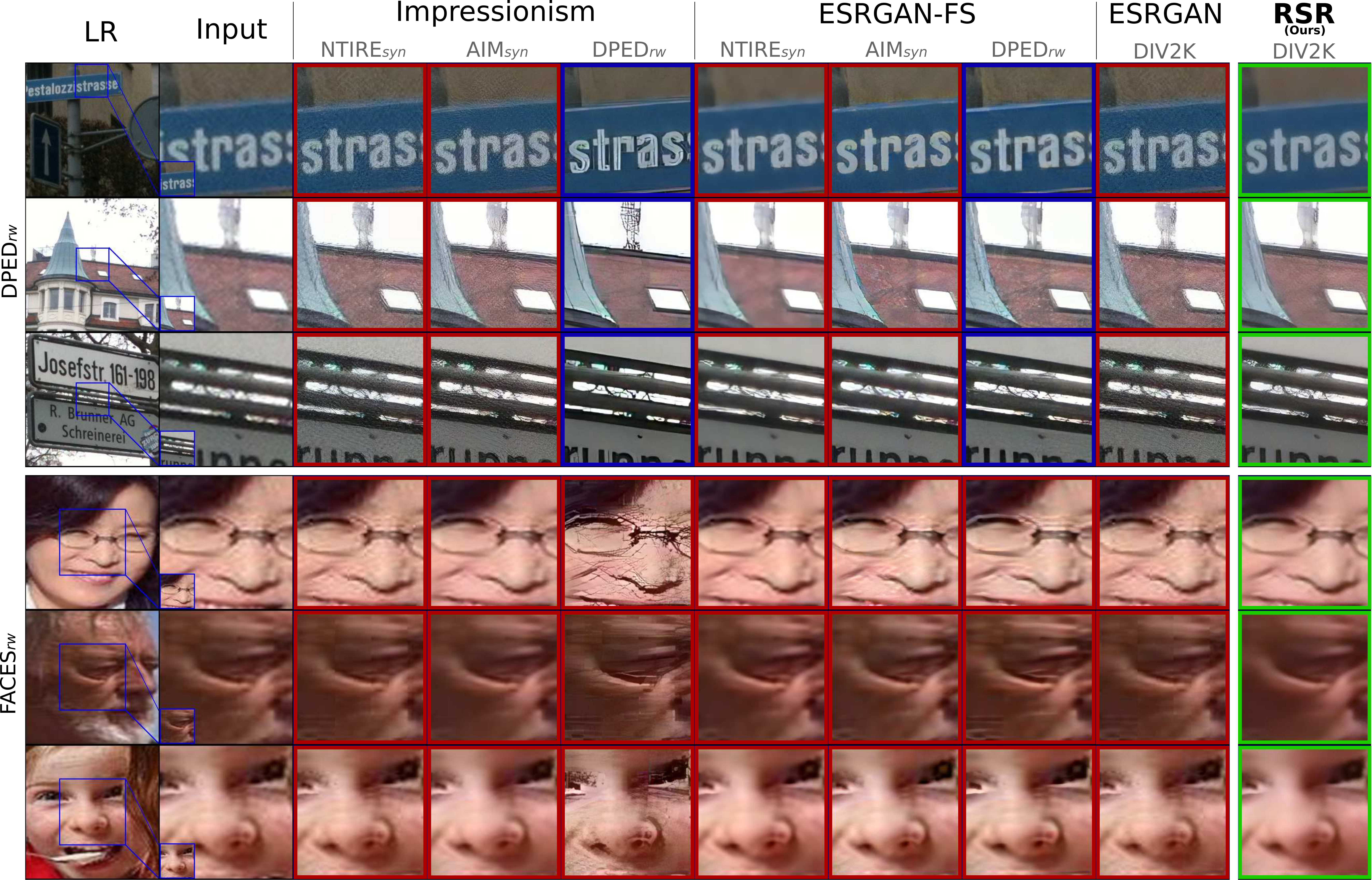}
  \caption{\textbf{Qualitative comparison on real-world images.} Comparison between our method and state-of-the-art methods, for two real-world datasets: \DPED~and \FACES. 
  Note that these datasets do not have ground-truth.
  {\color{blue}Blue} frames denote training and validation on the same dataset. {\color{red}Red} frames denote training and validation on different datasets. {\color{green}Green} frames denote our method.}
  \label{fig:results_rwsr}
\end{figure*}

\subsection{Real-World Synthetic Datasets} 
The winners of the two recent real-world SR challenges employ noise-specific models, \ie, different models depending on the type of corruption. To compute their generalization capabilities, in \tref{tab:synthetic_images}, we evaluate each specialized model on both synthetic datasets \NTIRE~and \AIM~and rank them by their average performance. State-of-the-art models perform well on the dataset they were trained on but have lower performance on unseen corruption statistics. Given that we use a \emph{single} model and it is not statistically-dependent on the type of artifacts, our generalization capabilities are considerably better than those of previous works. \fref{fig:result_paired} further reinforces this notion by showing the artifacts that state-of-the-art models create on unseen datasets (red frames on the Fig.). See the supplementary material for an in-depth study of the type of artifacts that each training dataset enforces. In comparison, we have a single robust model that achieves the best LPIPS for both unseen datasets and successfully removes input noise without creating artifacts. 

\subsection{Real-World Datasets} 
\fref{fig:results_rwsr} shows the output of different methods for two challenging in-the-wild datasets: \DPED~\cite{ignatov2017dslr} and \FACES~\cite{yang2016wider}. A traditional SISR method~\cite{wang2018esrgan} and specialized real-world models lack removing unseen real-world noise. Note that Impressionism trained on \DPED~creates sharp super-resolved images on the same dataset (blue frames), but produces less pleasant images with stronger artifacts in \FACES. This result suggests that, even with a real-world noise training, 
there is no guarantee the model generalizes well towards images in the wild. Conversely, our model removes real-world noise from both datasets, having better performance on input images with clear edges (\eg, street sign from \DPED). Thus, we show that our \emph{single} robust model accurately super-resolves unseen synthetic and natural real-world images. 

\section{Conclusion}
In this work, we explore the use of adversarial attacks to improve the robustness to unseen noise in the task of real-world SR. We present RSR, a novel SR method that leverages robust adversarial examples to create photo-realistic HR images regardless of the LR input noise. We evaluate our method on synthetic and natural real-world SR datasets. By using a \textit{single} robust model trained only with a clean dataset, we outperform current state-of-the-art methods requiring specialized models for each type of corruption. Furthermore, we provide theoretical insights on how to adapt adversarial attacks for the particular needs of real-world SR models. We expect our work to catalyze further study of the fruitful relationship between adversarial robustness and real-world SR. 

\noindent \textbf{Acknowledgements:} We thank Guillaume Jeanneret for insightful discussions on the subject.  

{\small
\bibliographystyle{ieee_fullname}
\bibliography{egbib}
}

\end{document}